\documentclass[a4paper,11pt]{article}
\pdfoutput=1 

\usepackage{jinstpub} 

\usepackage{lineno}
\usepackage{graphicx}
\usepackage{textcomp}

\newcommand {\idark}       {\mbox{I$_{\rm dark}$ }}

\title{\boldmath An induced annealing technique for SiPMs neutron radiation damage}

\author[a]{M.~Cordelli}
\author[a]{E.~Diociaiuti}
\author[c]{A.~Ferrari}
\author[a]{S.~Miscetti}
\author[c]{S.~M\"uller}
\author[b]{G. Pezzullo}
\author[a,1]{I. Sarra\note{Corresponding author.}}

\affiliation[a]{Laboratori Nazionali dell'INFN, via Enrico Fermi 54, 00044 Frascati, Italy}
\affiliation[b]{Department of Physics, Yale University, 56 Hillhouse, New Haven, CT-06511, USA}
\affiliation[c]{Helmholtz-Zentrum Dresden-Rossendorf (HZDR), Bautzner Landstra{\ss}e 400, 01328 Dresden, Germany}
\emailAdd{ivano.sarra@lnf.infn.it}

\abstract{
The use of Silicon Photo-Multipliers (SiPMs) has become popular in the
design of High Energy Physics experimental apparatus with a growing
interest for their application in detector area where a significant
amount of non-ionising dose is delivered. For these devices, the main
effect caused by the neutron fluence is a linear increase of the leakage
current.\\ In this paper, we present a technique that provides a
partial recovery of the neutron damage on SiPMs by means of an
Electrical Induced Annealing. Tests were performed, at the temperature of 20\textdegree{}C, on a sample of
three SiPM arrays ($2\times3$) of 6 mm$^2$ cells with $50\ \rm \mu m$
pixel sizes: two from Hamamatsu and one from
SensL. These SiPMs have been exposed to neutrons generated by the Elbe Positron Source facility (Dresden), 
up to a total fluence of $8 \times 10^{11}$ n$_{\rm 1
  MeV-eq}$/cm$^2$.  Our techniques allowed to reduced the leakage
current of a factor ranging between 15-20 depending on the overbias
used and the SiPM vendor. Because, during the process the
SiPM current can reach O(100 mA), the sensors need to be operated in a condition that
provides thermal dissipation. Indeed, caution must be used when applying this kind of 
procedures on the SiPMs, because it may damage permanently the devices themself.
}


\begin{document}
\maketitle
\flushbottom

\section{Introduction}
A Silicon Photo-Multiplier (SiPM) is a novel semiconductor
photo-detector composed by a matrix of pixels operating few volts
above the breakdown voltage (Geiger-Mode Avalanche
Photo-Diode). Distinctive features of such technology are: single
photon detection capability, high gain $\sim$ 10$^6$ and good time
resolution ($<$ 1 ns)~\cite{MPPCHAMAMATSU}.  In addition SiPMs present
small size and customisable granularity, insensitivity to magnetic
fields and are relatively un-expensive.  For these reasons several
current and planned High Energy Physics experiments employ SiPMs in
their experimental setup ~\cite{MU2ETDR,CMSHCAL}. A growing interest
exists in their application when a high level of non-ionizing dose is
expected to be delivered to the SiPMs.\\ Different studies showed a
correlation between the bulk defects in the Silicon structure due to
the radiation damage and the deterioration of the photo-detector
performances~\cite{1748-0221-12-07-C07030,DisplacementDamageEffects}.
Hadrons and high energy leptons can produce point defects as well
as cluster related defects in the photo-detector active volume. In
particular, neutrons travelling within the Silicon lattice induce many
displacements of Silicon atoms that, at the end of the path, form a
disordered agglomeration of atoms called
cluster~\cite{RadiationInducedPointandClusterRelated}. From the
macroscopic point of view, some of these defects act as charge
carriers generator centers, producing an increase of dark noise.  
The current state of the art technology does not provide any method to fully recover neutron
induced damage in Silicon based photo-detectors. Thermal annealing can
partially recover the neutron induced damage, as shown in
Ref.~\cite{Garutti}, where after several days of thermal
treatment at different temperatures, the dark current got reduced by a factor of about 10. 
Our tests reported in the next sections indicate that an important
recovery of the SiPM damage can be achieved also by means of an electrical
induced annealing in few minutes of treatment. Two different methods have been tested: a direct
and an inverse polarization of the SiPM using a non conventional over
voltage - larger than the usual 5~V limit indicated by several
vendors~\cite{MPPCHAMAMATSU,SENSLSIPM}. We believe that this technique may combine two effects: 
the thermal annealing induced by heating the Silicon lattice
and the presence of a strong electric field to re-order the atoms in
the Silicon lattice displaced by the neutron-induced bulk
damage~\cite{RANCOITA}.

\section{Experimental setup}
\begin{figure}[h!]
  \centering
  \includegraphics[width=0.76\textwidth] {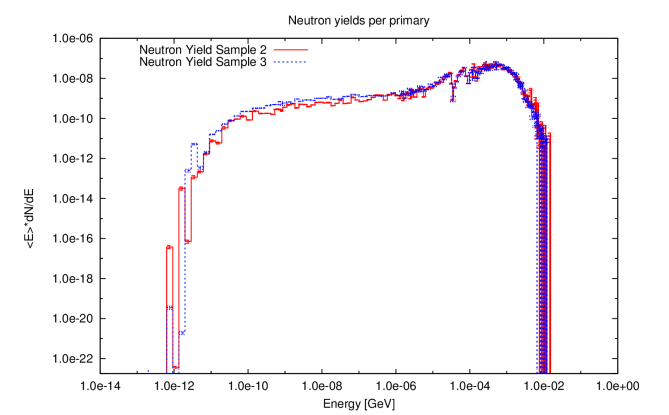}
  \caption{Fluka simulation of the neutrons energy spectrum at the SiPMs position.}
  \label{fig:neutronsMeV}
\end{figure}

The measurements described in this paper were performed at the Silicon
Detector Facility of the Fermi National accelerator
Laboratory~\footnote{http://www.fnal.gov}. Two different sets of SiPM
arrays were studied: Hamamatsu~\cite{MPPCHAMAMATSU} and
SensL~\cite{SENSLSIPM}. A sample from each vendor has been exposed to neutrons generated by the Elbe Positron Source facility (Dresden), up to a total fluence 
of $8 \times 10^{11}$ n$_{\rm 1 MeV-eq}$/cm$^2$~\cite{Cordelli:2017dgl}. The average energy of neutrons produced has
been estimated by a full FLUKA simulation; the spectrum, Figure~\ref{fig:neutronsMeV}, is well centred around 1 MeV.

\begin{figure}[h!]
  \centering
  \includegraphics[width=0.5\textwidth] {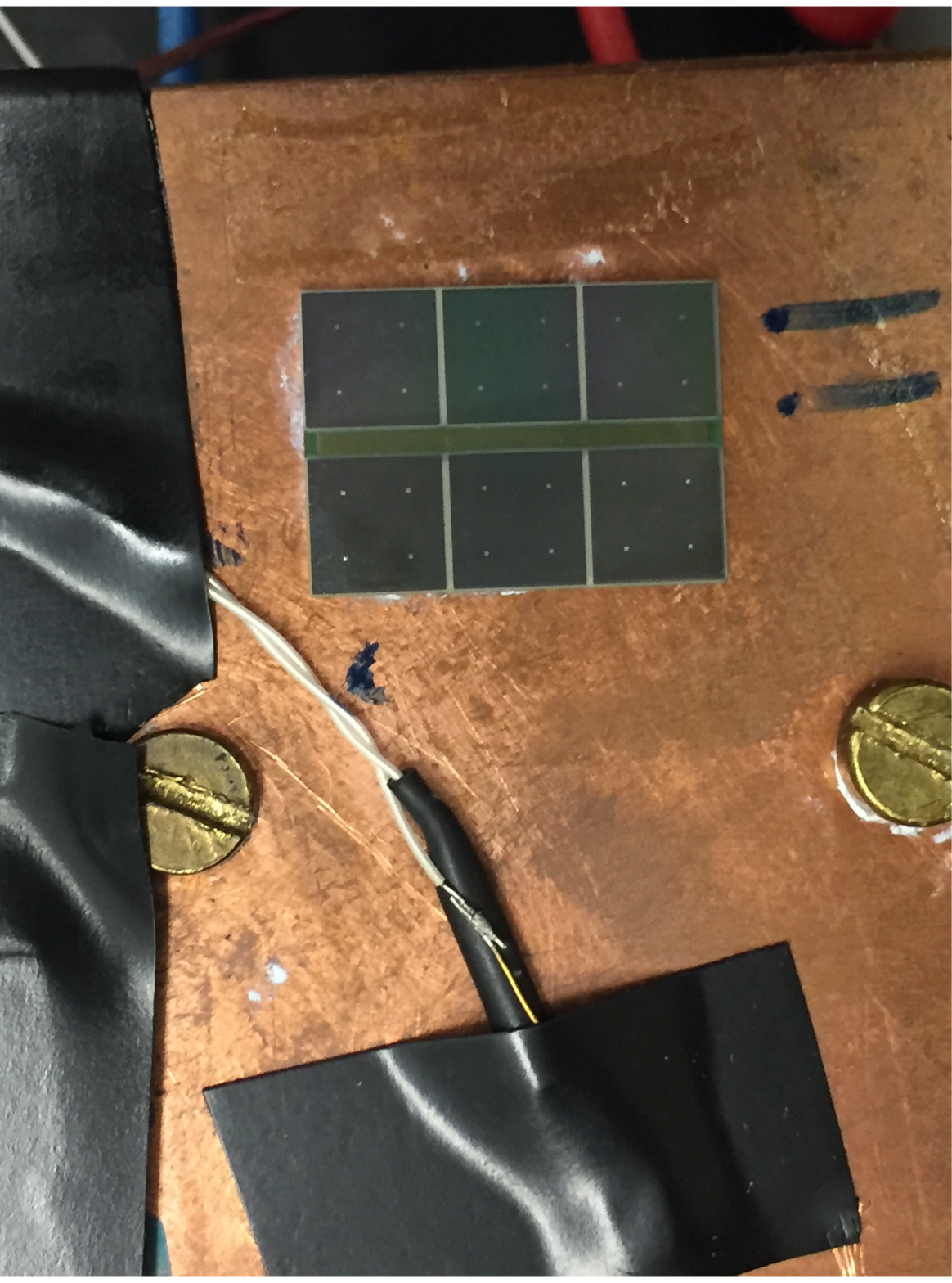}
  \caption{Cu support with one PT1000 and one SiPM array plugged in.}
  \label{fig:setup}
\end{figure}

The tests described in this paper has been conducted after 3 months from the irradiation. During most of the time, the SiPMs were kept 
at room temperature of about 20~\textdegree{}C and in a dark box.\\
These SiPMs consist of a 2×3 array of 6×6 mm$^2$ monolithic cells with pixel sizes of 50 $\mu$m. Both SiPM arrays have a custom design,
developed for the electromagnetic calorimeter of the Mu2e experiment~\cite{BIODOLA},~\cite{IVANO}.  
To improve the thermal dissipation capabilities compared to commercial designs\footnote{Hamamastu and SensL, private
communication}, they have been built with a thermal resistance of about $5\times 10^{-4}\;\rm{m^2K/W}$ . 
The experimental setup consisted of:
\begin{itemize}
\item
  light tight thermal chamber TestEquity Model
  140~\cite{TEMP_CHAMBER};
\item
  power supply PLH250~\cite{PLH250} to bias the SiPMs
  and measure their current;
\item
  $30\times20$ $\rm cm^2$ Cu plate, 3 mm thick, used as thermal
  support for the SiPMs;
\item
  3 PT1000~\cite{PT1000} thermistors used to measure the temperatures
  of the SiPM and of the Cu support.
\end{itemize}
Figure~\ref{fig:setup} shows a picture of the Cu support with the PT1000, used 
to measure the temperature of the support, and one SiPM array plugged in. 
All the tests were performed inside a thermal chamber at a temperature of 20~\textdegree{}C.
This setup allowed us to monitor the temperature of the SiPM active
region and of the Cu support during all the measurements and bias the SiPM
with a current limitation of few hundreds mA. A second PT1000 sensor was placed on the top of the silicon resin of the 
SiPMs; the third PT1000 sensor was used to check that the temperature drop on the pin-side of the
SiPM was within 2~\textdegree{}C with respect to the one measured on
the active region.

\section{SiPM recovery measurements}
Two different configurations for biasing the SiPMs were tested: direct
and inverse polarization using an over-voltage larger than 5 V.\\ So, we
define:
\begin{itemize}
\item
 $\rm V_{DIR}$ as the direct voltage, positive voltage applied between anode and cathode;
 \item 
 $\rm \Delta V_{IND}$ as the reverse over voltage with respect to the breakdown voltage, $\rm V_{br}$, positive voltage applied between cathode and anode;
 \item 
 $\rm I_{TEST}$ as the photo-detector current during the annealing procedures;
 \item 
 \idark as the leakage current measured in the standard reverse configuration of the SiPM at the operation voltage $\rm V_{op} = V_{br}$ + 3 V. 
 \end{itemize}
 
 All the results shown in the following are relative to the \idark at this operational voltage point.  At each annealing step, the determination of \idark has been done 
 stopping the annealing, carrying the device at 20~\textdegree{}C with the thermal chamber and then measuring the current. 

\subsection{Induced annealing in direct polarization}
\begin{figure}[h!]
  \centering
  \includegraphics[width=0.76\textwidth] {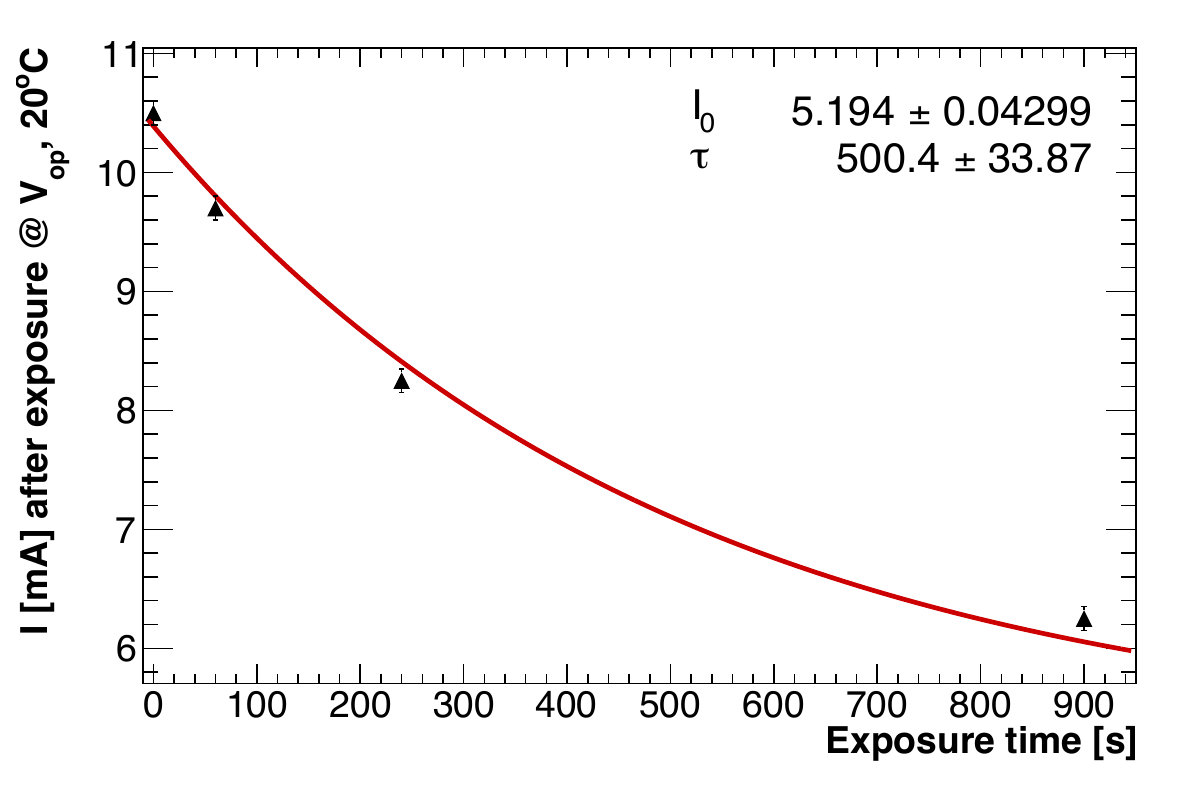}
  \caption{\idark as a function of the exposure time for a direct polaritation. Points were fit
    with an exponential curve.}
  \label{fig:direct_polarization}
\end{figure}

Figure~\ref{fig:direct_polarization} shows the results obtained with
one cell of a Hamamatsu SiPM array polarized in direct configuration
at $\rm V_{DIR}$ = 10 V. The \idark shows an exponential trend
with a decay time of $\tau=500$~s and a constant current term $\rm
I_{0 dark}$ = 5.2~$\rm m A$\footnote{Fit function used: $\rm{I_{dark}=I_{0 dark}\cdot(1+exp(-t/\tau))}$.}. A reduction of $\sim40\%$ in \idark after
$\sim 5$~min of direct polarization was achieved.
In this configuration the SiPM was generating a $\rm I_{TEST}$ of $\sim530$
mA, corresponding to a power of $\sim 5$ W, that is the power needed to recover the damage of a factor of about 2. 
\subsection{Induced annealing in reverse polarization}
\begin{figure}[h!]
  \centering
  \includegraphics[width=0.76\textwidth] {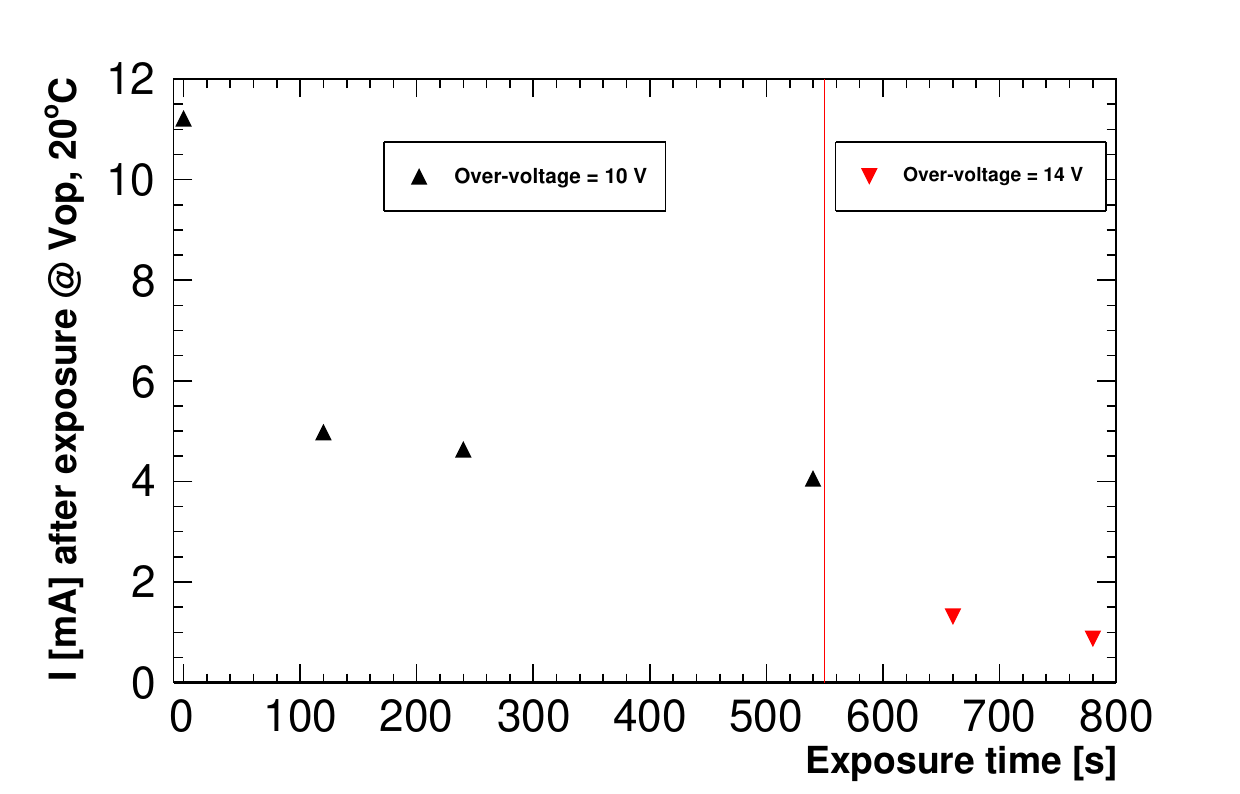}
  \caption{I$_{\rm dark}$ as a function of the exposure time for a
    Hamamatsu SiPM cell for an inverse polarization. The two different settings for
    the over-voltage used during the test are separated by the vertical red line and indicated in the legends.}
  \label{fig:dresda_cell}
\end{figure}

\begin{figure}[h!]
  \centering
  \includegraphics[width=0.76\textwidth] {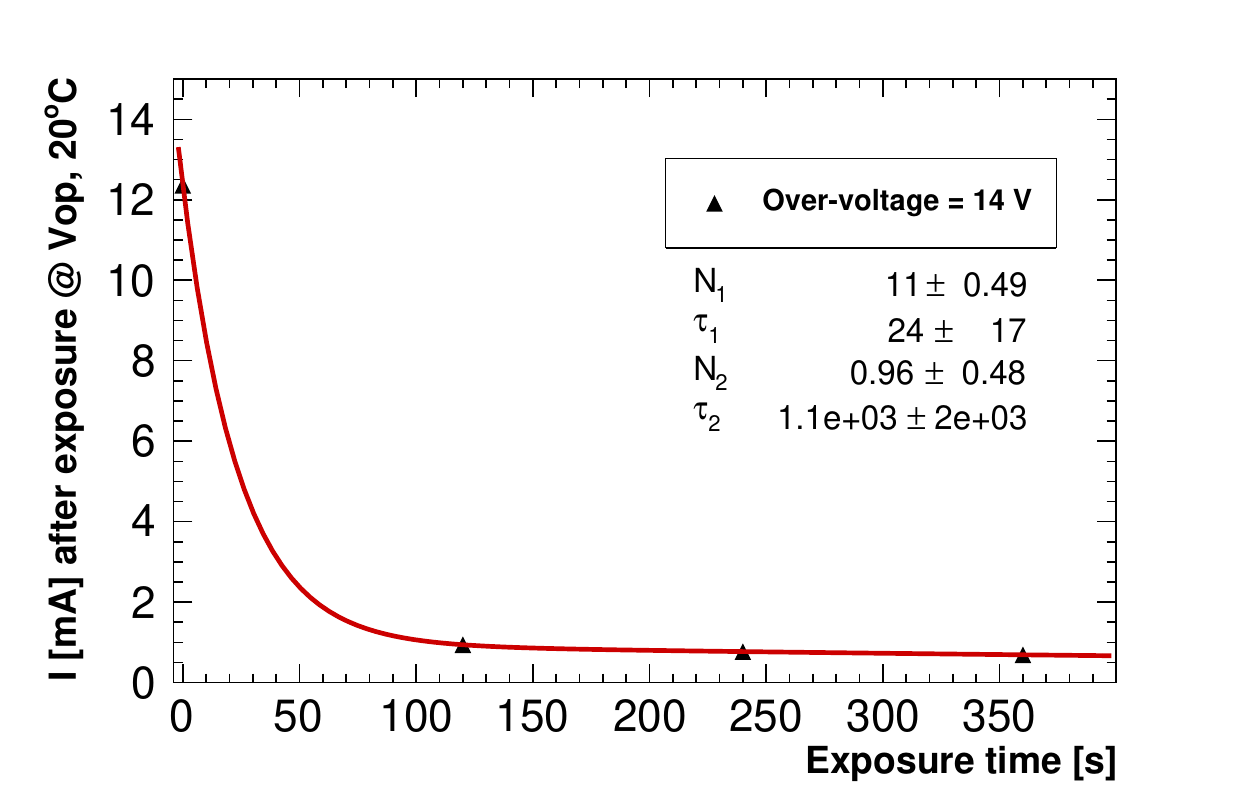}
  \caption{\idark as a function of the exposure time for a Hamamatsu
    SiPM cell applying an over-voltage of 14~V during the recovery.}
  \label{fig:dresda_cell_1a}
\end{figure}

To quote the performance of the recovery method in inverse polarization 
two different settings of over voltage were tested. 
Both Hamamatsu and SensL
SiPMs have been checked in this configuration. Figure~\ref{fig:dresda_cell} shows
the \idark variation as a function of the exposure time for a
Hamamatsu SiPM. The first four measurements were taken at $\rm \Delta V_{IND}$ = 10 V, 
then we improved the
recovery by increasing the over voltage up to $\rm \Delta V_{IND}$ = 14 V.
After about 13 min at the described conditions,
the SiPM \idark reached about $\rm 1$ mA that corresponds to a
reduction of a factor of 10 when compared to the starting
value. Finally, we tried to increase the over-voltage up to $\rm \Delta V_{IND}$ = 17 V, but at
this value the SiPM got broken.
The $\rm I_{TEST}$ measured during this operation was 130 (186) mA for
a $\rm \Delta V_{IND}$ of 10 (14) V, corresponding to a dissipated
power of about 8 (12)~W. The temperature measured on the SiPM active
region with the PT1000 was $\sim 150$~\textdegree{}C.\\ After the learning phase, we
tested another Hamamatsu SiPM cell of the same device setting $\rm
\Delta V_{IND}$ at 14 V. Figure~\ref{fig:dresda_cell_1a} shows the
results of the \idark measurement operated every 2 min; from the fit\footnote{Fit function used: $\rm{I_{dark} = N_1 \cdot exp(-t/\tau_1) + N_2 \cdot exp(-t/\tau_2)}$.} 
we observe that a rapid reduction of the leakage current occurs with a $\tau$ of the order of 25 seconds . After about 6
min of electrical annealing, a total reduction of a factor 15 on the
leakage current was observed, thus indicating a good agreement with the previous test done on the same 
SiPM model.

\begin{figure}[h!]
  \centering
  \includegraphics[width=0.76\textwidth] {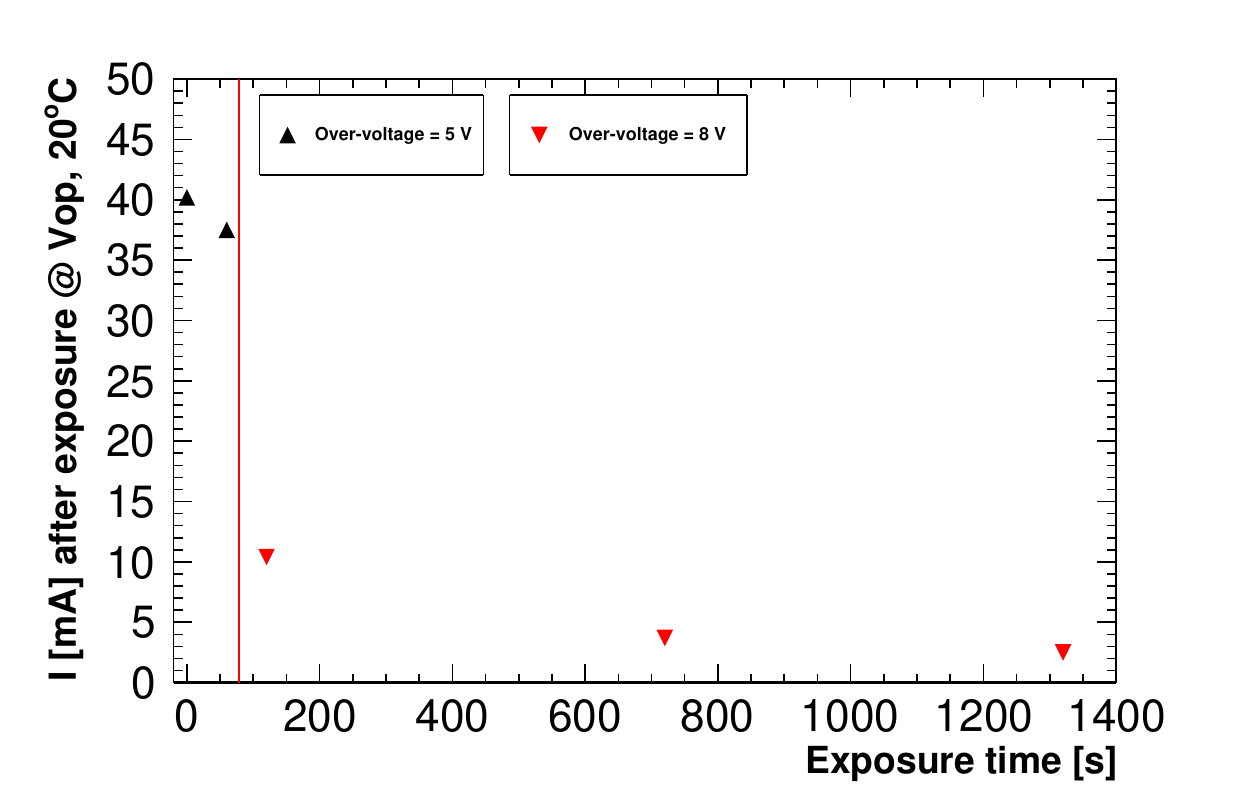}
  \caption{I$_{\rm dark}$ as a function of the exposure time for a
    SensL SiPM cell. The two different settings for
    the over-voltage used during the test are separated by the vertical red line and indicated in the legends}
  \label{fig:sensl_cell_scan}
\end{figure}

\begin{figure}[h!]
  \centering
  \includegraphics[width=0.55\textwidth,angle=-90] {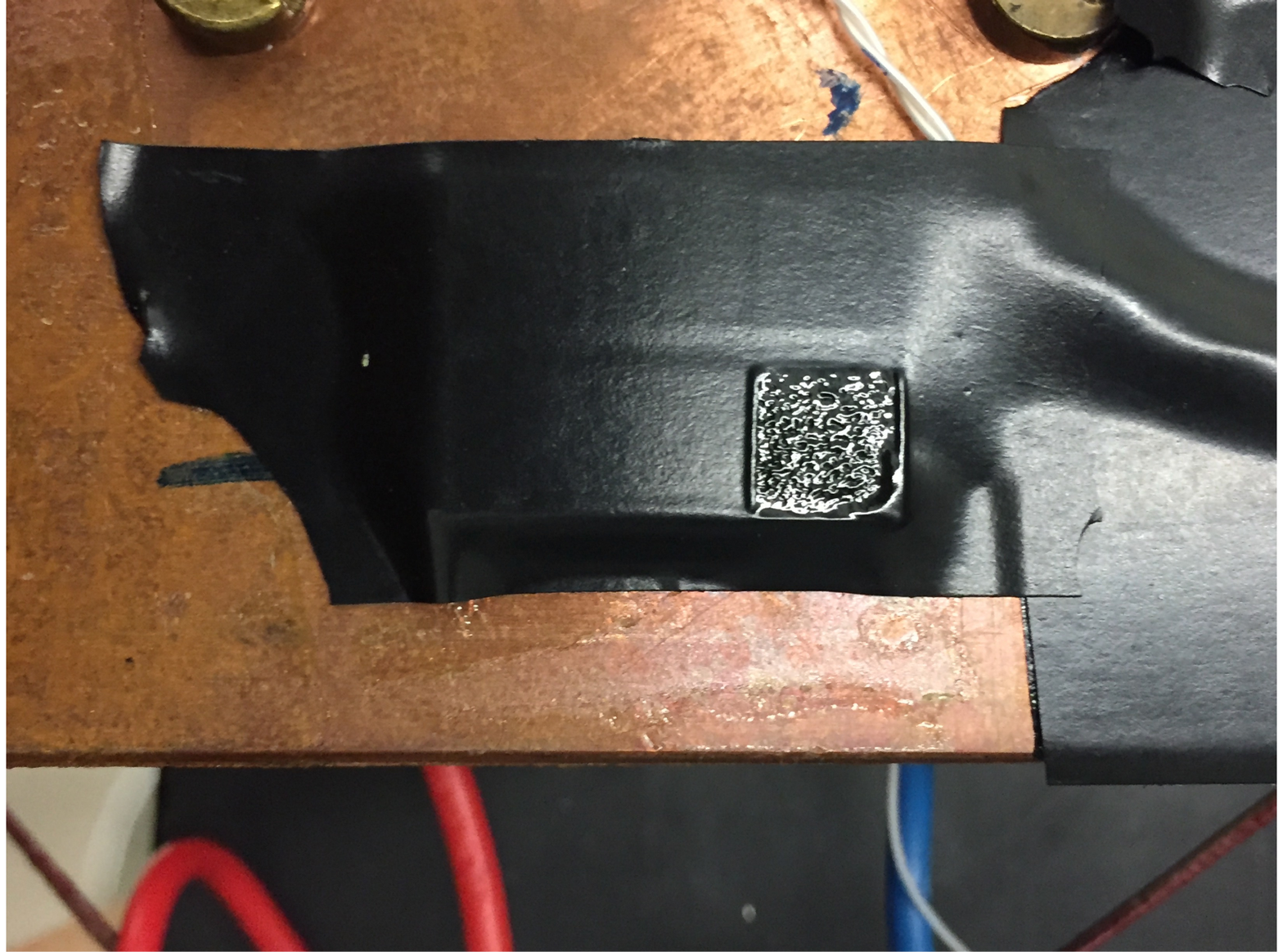}
  \caption{SiPM array on the Cu support at the end of the tests.}
  \label{fig:sipm_heat}
\end{figure}

\begin{figure}[h!]
  \centering
  \includegraphics[width=0.76\textwidth] {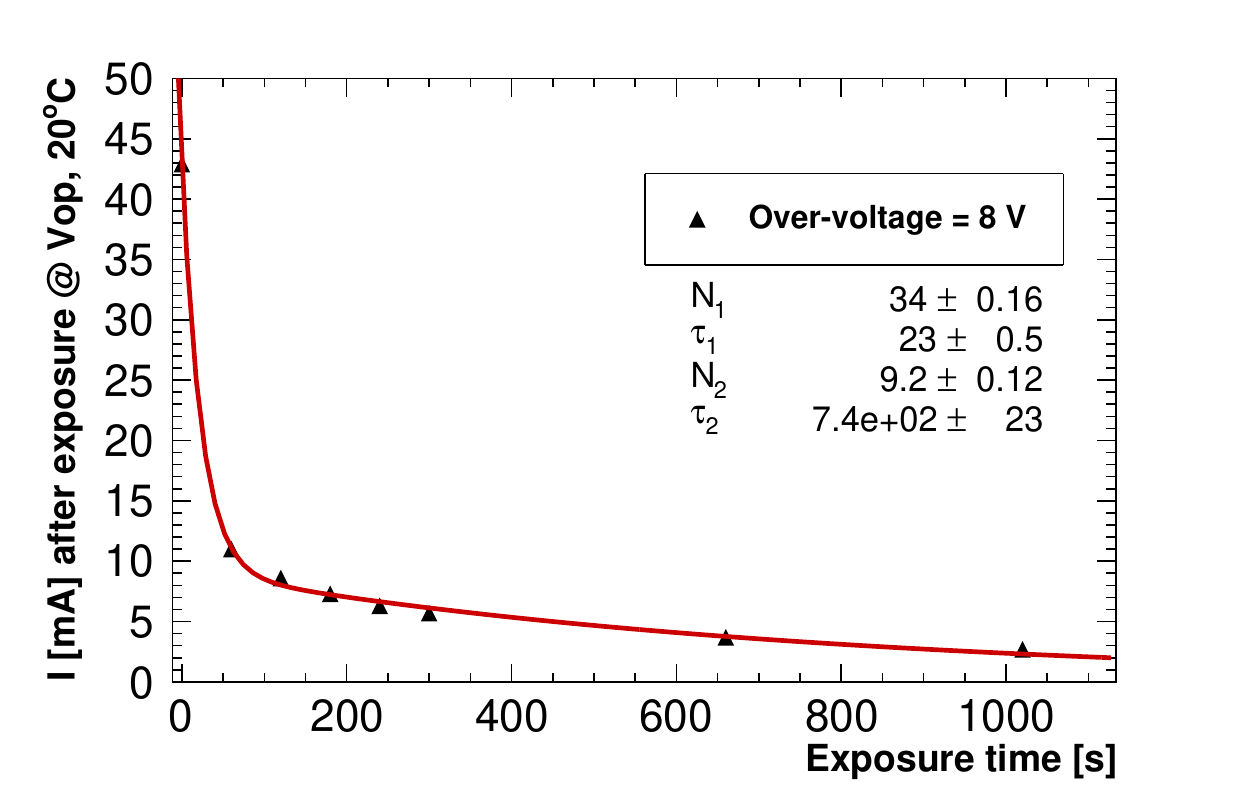}
  \caption{I$_{\rm dark}$ as a function of the exposure time for a
    SensL SiPM cell applying an over-voltage of 8~V during the recovery.}
  \label{fig:sensl_cell}
\end{figure}

A similar test has been carried out also for the SensL
device~\cite{SENSLSIPM}. Figure~\ref{fig:sensl_cell_scan} shows the
measured \idark for two different settings of $\rm \Delta
V_{IND}$\footnote{The breakdown voltage of the SensL SiPM at
  20~\textdegree{}C is 24.87 V~\cite{SENSLSIPM}.}: 5 and 8 V. After 22
min we measured on overall reduction factor $\sim16$ on \idark (from
40.2 to 2.5 mA).
Unfortunately after 22 min the SiPM got unplugged from the Cu plate,
reached a temperature larger than 200~\textdegree{}C and got
broken. Figure~\ref{fig:sipm_heat} shows the localization of the
heating induced by the over voltage condition on the single cell under
bias.
During the annealing treatment, we have measured a $\rm I_{TEST}$
current of 200 (400) mA  for $\rm \Delta V_{IND}$ of 5 (8)
V, corresponding to a dissipated power of about 6 (13)~W. The
temperature measured on the SiPM active region was $\sim
150$~\textdegree{}C.\\
We repeated the measurement on another cell of the same device
at $\rm \Delta V_{IND}$ = 8 V. Figure~\ref{fig:sensl_cell} shows the \idark
variation as a function of the exposure time; also in this case,
we observe that a rapid reduction of the leakage current occurs with a $\tau$ of the order of 25 seconds, consistent with the value observed in Figure \ref{fig:dresda_cell_1a}. After
$\sim 17$ min the \idark was reduced from 42.9 to 2.7 mA that
is compatible with the reduction observed in the previous test.

\section{Comparison between induced and thermal annealing}
To understand if the observed reduction of \idark was due only to the
related thermal annealing, we exposed the SiPM to high temperatures at
$\rm V_{op}$. Since the breakdown voltage of SiPMs changes in an inversely proportional way 
with the temperature following a rule of about 0.1\%/\textdegree{}C, during the test we have adjusted 
the breakdown voltage with respect to the temperature of the test. For this test we used a different cell of
the Hamamatsu SiPM array. We started keeping the SiPM at
80~\textdegree{}C for 19 min, then increased the temperature up to
120~\textdegree{}C, following the procedures explained in ~\cite{Garutti}. Results are summarised in Table~\ref{table:thermal_test}. 
\\
\begin{table}[h!]
  \begin{center}
    \begin{tabular}{|c|c|c|} \hline
      {\bf SiPM T [$^{o}$C]} & {\bf Exposure time [s] } & {\bf I$_{\rm dark}$ [mA] }\\\hline
      20 $\pm$ 0.5  &    0 &  12.33 $\pm$ 0.01 \\\hline
      80  $\pm$ 0.5 & 1120 &   9.93 $\pm$ 0.01 \\\hline
      120 $\pm$ 0.5 &  600 &   9.50 $\pm$ 0.01 \\\hline
    \end{tabular}
  \end{center}
  \caption{I$_{\rm dark}$ measured @ 54.7 V, 20$^{o}$C after
    exposure of the SiPM at different thermal conditions.}
  \label{table:thermal_test}
\end{table}

In 30 minutes, the thermal annealing alone provided a recovery
 of about 25\% that is much smaller than the one observed
with electrical induced annealing in a much shorter time.

\section{Conclusions}
We have presented an electrical induced annealing technique that
allows to partially recover neutron damage of SiPM. Tests of such a
technique were performed on two different sets of SiPM arrays from
Hamamatsu and SensL. The benefit of the observed method is that it
allows to partially recover (up to a factor 15-20) the bulk damage in
the SiPM exposed to a neutron fluence of $8\times10^{11}\rm n_{\rm 1
MeV-eq}/cm^2$, thus improving
the results one may get with the conventional thermal annealing. A remarkable recovery related 
to high temperature annealing is presented in \cite{Tsang}, where 
results of few days of annealing at +250~\textdegree{}C, using
forward bias with the SiPM current reaching 10 mA, are shown. 
By comparison, the biggest advantage of our technique resides in the possibility to be used in an installed detector “in situ” with an application of few minutes.
In particular, we have observed that a rapid reduction of the leakage current occurs with a $\tau$ of the order of 25 seconds.\\
However, since during this process the
SiPM current can reach O(100 mA), the sensors need to be operated in a condition that
provides thermal dissipation. This precaution will avoid the sensors heating up to breaking 
temperatures or damage permanently the device hosting the sensors.

\acknowledgments{
This work was supported by the EU Horizon 2020 Research and Innovation Program under the
Marie Sklodowska-Curie Grant Agreement No. 690385. The authors are grateful to many people for
the successful realisation of the tests. In particular, we thank Dr. Adam Para, for providing us a
necessary and well equipped facility at SiDet, Fermilab.
}


\end{document}